\documentclass{article}

\usepackage{arxiv}
\usepackage{caption}
\usepackage{subcaption}
\usepackage{amsmath}
\usepackage[utf8]{inputenc} % allow utf-8 input
\usepackage[T1]{fontenc}    % use 8-bit T1 fonts
\usepackage{hyperref}       % hyperlinks
\usepackage{url}            % simple URL typesetting
\usepackage{booktabs}       % professional-quality tables
\usepackage{amsfonts}       % blackboard math symbols
\usepackage{nicefrac}       % compact symbols for 1/2, etc.
\usepackage{microtype}      % microtypography
\usepackage{lipsum}
\usepackage{graphicx}
\usepackage{xcolor}

\usepackage[backend=biber, sorting=none, maxbibnames=2,style=numeric-comp]{biblatex}
\graphicspath{ {./img/} }

\newcommand{\DistribFunc}{P}
\newcommand{\DistribAlphaBetaFunc}{\rho}
\newcommand{\FFunc}{F}

\title{Coarse-graining effect supports EDR in axonal wiring databases}

\author{
 Máté Józsa \\
  Department of Physics\\
  Babeş-Bolyai University\\
  Str. M. Kogalniceanu 1, 400084 Cluj-Napoca, Romania \\
  \texttt{mate.jozsa@ubbcluj.ro} \\
   \And
 Mária Ercsey-Ravasz\\
  Department of Physics\\
  Babeş-Bolyai University\\
  Str. M. Kogalniceanu 1, 400084 Cluj-Napoca, Romania \\
  Transylvanian Institute of Neuroscience \\
  Network Science Lab\\
  \texttt{maria.ercsey@ubbcluj.ro} \\
  \And
 Zsolt I. Lázár\\
  Department of Physics\\
  Babeş-Bolyai University\\
  Str. M. Kogalniceanu 1, 400084 Cluj-Napoca, Romania \\
  Transylvanian Institute of Neuroscience \\
  Network Science Lab\\
  \texttt{zsolt.lazar@ubbcluj.ro} \\
}

\begin{document}
\maketitle
\begin{abstract}
The Exponential Distance Rule (EDR) is a well-documented phenomenon suggesting that the distribution of axonal lengths in the brain follows an exponential decay pattern.
Nevertheless, the availability of individual-level axon data supporting this assertion is limited, while inter-regional connectome data is accessible for several species, including Drosophila, mouse, macaque, and human. While, axon-level data in Drosophila and mouse support the generality of the EDR rule, with the possibility of determining the $\lambda$ decay rate as the inverse of the average axonal length, region-level data can vary significantly from the exponential curve.
In this study, we establish that the length distribution of region-level data converges onto a universal curve when rescaled to the mean axonal length, thereby demonstrating similarities across different species. 
To explain this, we present a simple mathematical model that accurately accounts for the observed deviations from the EDR in the weighted length distribution of inter-regional connectomes.
Our model takes into account the inherent effects of coarse-graining while carefully considering limitations, such as the influence of dimensionality, curvature, and the effect of resolution. 
We illustrate the robustness of our model through numerical simulations and by contrasting our hypothesis with data from a neuronal null model.
The findings validate the universality of the EDR rule across various species, paving the way for further in-depth exploration of this remarkably simple principle.

\end{abstract}

\keywords{EDR \and coarse-graining \and distance distribution \and axonal connections}

%%%%%%%%%%%%%%%%%%%%%%%%%%%%%%%%%%%%%%%%%%%%%%%%%%%%%%%%%%%%%%%%%%%%%%%%%%%%%%%%%%%%%%%%%%%
\section{Introduction}
%%%%%%%%%%%%%%%%%%%%%%%%%%%%%%%%%%%%%%%%%%%%%%%%%%%%%%%%%%%%%%%%%%%%%%%%%%%%%%%%%%%%%%%%%%%
The brain is acclaimed as one of the most complex biological systems. Not only its functioning but also its structure still holds a lot of open questions.
According to a simplified picture of the structure of the neocortex, also known as the grey matter, we can view it as a system made of an immense number of neurons, cells specialized in non-local directed interactions facilitated by their particular shape characterized by extreme aspect ratios.
The number of connections a neuron maintains with other neurons can be in the order of thousands depending on the species.
From the perspective of network theory, the currently most widely accepted approach to complex systems, the brain can be regarded as a large directed network of neural soma connected by axons and synapses.
Within the context of structural network models the role of edges is played by the axons whereas synapses fulfill this function in frameworks concentrating on functional connectivity.
The synaptic connectome that underlies brain physiology exhibits a significantly more intricate topology in comparison to the axonal connectome, which tends to prioritize wiring cost optimization \cite{Bullmore2012}. This study will focus on investigating the latter.
The first species for which the complete neural system has been mapped was the ringworm C. elegans consisting of 302 nodes and around 5.000 edges  \cite{Cook2019}. 
Recently a detailed mapping of the neural system of the Drosophila (fruit fly) has been reconstructed with high fidelity using electron microscopy \cite{Scheffer2020}. 
The central brain region of Drosophila has approximately 32.000 nodes and 850.000  edges (see the upper part of Figure \ref{fig:dataPanel}a).
For larger animals, especially mammals mapping the whole brain on the neuronal level is not imaginable in the near future.
The human brain, for example, contains approximately 86  billion neurons with an average number of synaptic connections in the order of thousands \cite{Cook2019, CodexApp}. 
In these cases, connectomics usually deals with mapping the connections or axon bundles between different brain regions, such as functional areas.

There has been an ongoing effort to reconstruct these structural brain connectomes using a variety of noninvasive and invasive experimental methods \cite{Zeng2018}.
Non-invasive methods, such as diffusion tensor imaging, allow for the reconstruction of the brain's white matter tracts \cite{Sotiropoulos2017}. 
However, the method is not very precise in measuring the weight of connections between areas, and the directions of links also can not be determined.  
Invasive methods, such as retrograde and anterograde tract tracing, involve the injection of fluorescent substances into well-defined brain areas, allowing for the tracing of specific neural pathways (see the upper part of Figure \ref{fig:dataPanel}b) \cite{Lanciego2011}.
The exponential distance rule (EDR) was first observed in a macaque region-level dataset obtained by retrograde tracing experiments \cite{Markov2012,ErcseyRavasz2013} and later also confirmed by axon-level measurements in the mouse brain \cite{Horvt2016}. 

Recently, a comprehensive database at the neuron level for the Drosophila brain has been made public \cite{Scheffer2020}. Data pertaining to the central brain region, extracted from this database, appears to further support the EDR.  	
According to this rule the number of axons crossing the white matter decreases exponentially with their length, i.e., the normalized distribution of axonal lengths shows an exponential decay (see the bottom panel of Figure \ref{fig:dataPanel}a) \cite{Horvt2016}: 
\begin{equation}\label{E:exponential}
\DistribAlphaBetaFunc_\lambda(s)=\lambda \exp(-\lambda s)~.
\end{equation}
One can easily show from this that the decay rate in the EDR is the inverse of the average axonal length: $\lambda=1/ \langle s\rangle$.
When the distances are normalized by the average length and $\langle s\rangle \DistribAlphaBetaFunc_\lambda(s)$ is plotted against $s/ \langle s\rangle$, the data curves from various species are expected to converge on an exponential curve with a decay rate of one.

Based on this phenomenon a simple network model, the EDR model was built, and has been shown to explain a number of topological properties of the brain connectome  \cite{ErcseyRavasz2013,Horvt2016,Gmnu2018,Wang2020}. 
This finding has sparked interest as, if confirmed, it indicates some simple principles underlying the macroscopical organization of the brain \cite{Song2021}. 
On one side, the EDR is expected to be the result of geometrical constraints on the connections within this physical network\cite{Bassett2018, Posfai2023}. 
On the other hand, it also illustrates wiring optimization in the brain, characterized by a large number of short connections and a minimal number of long connections, resulting in efficient communication in terms of material costs \cite{Markov2013}. 
This rule finds additional support in a recent study on the physical nature of spatial networks, offering an analytical framework to explore physically constrained jammed-state networks, such as the connectome \cite{Posfai2023}.

However, another commonly used class of structural connectivity measurement techniques does not differentiate between the lengths of each individual axon. 
Instead, these techniques provide the number of axons crossing between a specific pair of brain regions and the average length of these axons, as measured along the axon bundles connecting these two regions.\cite{Betzel2018, Oh2014, Chiang2011, Markov2013, Cammoun2012}.

The distance rule, extracted from the region-level connectomes, exhibits a high degree of universality across species when distributions are normalized to the mean (see bottom panel of Figure \ref{fig:dataPanel}b). 
Though not qualitatively different, there is a discernible deviation from the EDR pattern observed for  small distances,  and also at longer ones in the slope of the distribution.
%Figure \ref{fig:dataPanel} offers a visual comparison, elucidating these distinctions.
This apparent deviation from the exponential pattern in region-level data suggests a different manifestation of the EDR, compared to neuron-level observations. 

Our hypothesis asserts that the EDR accurately describes neuron connectivity, as evidenced by the detailed data from mice and Drosophila at the neuron level. 
We propose that the differences observed at the region level are not inconsistencies, but rather an alternate manifestation of the EDR, resulting from the aggregation of individual neuron endpoints into single points for each brain region. 
This aggregation reduces the variability found in neuron-level data. Our aim is to demonstrate that region-level data is consistent with the neuron-level EDR.
We aim to demonstrate that experimental data from various regions, spanning from Drosophila to humans, converge onto a universal master curve when normalized to the mean axonal length.
The form of this curve can be effectively described by a straightforward analytical model.
A strong indication for the coarse-graining effect is the significant reduction in the probability for small distances and in the slope of the distribution (see bottom panels of Figure \ref{fig:dataPanel}) 
In the following sections, we first provide detailed insights into the experimental structural connectivity data mentioned earlier. 
Then, we introduce a straightforward one-dimensional model designed to reconcile the two types of data, neuron-level and region-level wiring distances. 
We scrutinize the opportunities and challenges associated with employing this oversimplified analytical solution and address the case of higher dimensions, curvature of space, and the effect of resolution.

%%%%%%%%%%%%%%%%%%%%%%%%%%%%%%%%%%%%%%%%%%%%%%%%%%%%%%%%%%%%%%%%%%%%%%%%%%%%%%%%%%%%%%%%%%%
\section{Data}\label{sec:data}
\begin{figure}[h!]
	\centering
	\includegraphics[width=\linewidth]{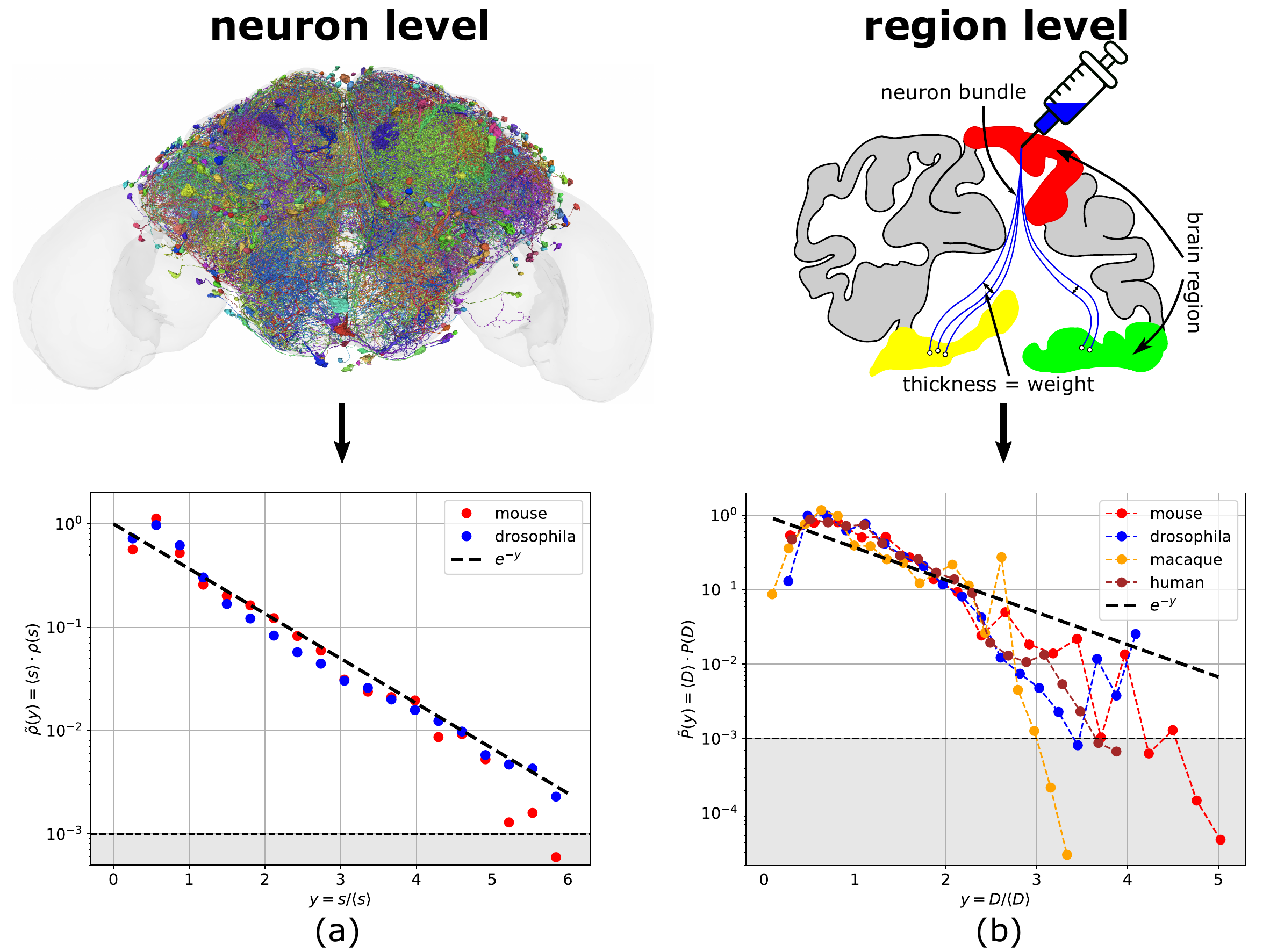}
	\caption{
		Illustrations depicting two distinct types of structural brain connectome data.
		In the first scenario (a), exhaustive neuron level data is presented. 
		The top left panel presents the central connectome of the Drosophila encompassing thousands of axons (reconstructed with Codex Web App \cite{CodexApp}).
		In this approach, the distribution of axonal lengths is formulated using information derived from individual neurons.
		In the lower-left panel, the distance distribution is depicted for a dataset comprising 1,984,074 individual neurons in the mouse brain, alongside data from the central connectome of the Drosophila fly (as illustrated in the upper panel) with a total of 32,422 cells \cite{Horvt2016, Scheffer2020}.
		These analyses constitute thorough examinations of axonal lengths that underlie the Exponential Distance Rule (EDR) \cite{Horvt2016}.
		In the second scenario (b), a coarse-grained (weighted) region-level approach is employed.
		This approach entails the application of tract-tracing techniques to the brain.
		In this context, brain regions are visually distinguished by distinct colors, and axonal bundles function as connections between these regions (see top-right panel).
		%The weight of a connection is determined by the number of axons within the corresponding bundle .
		In this case, the distribution of axonal lengths is computed by taking into account the lengths of the axonal bundles, with each bundle being weighted according to the number of axons it contains.
		The lower-right panel illustrates the distribution of weighted inter-region distances across multiple species, including macaque, Drosophila, mouse, and human \cite{Markov2012, Betzel2018, Chiang2011, Gmnu2018}.
		Discrepancies between the two types of data (lower-left and lower-right panels) become apparent when examining the distribution shape for short distances and observing the corresponding slope for bigger distances. 
		The dashed line on this figure represents the EDR based theoretical limit of infinitely high resolution, i.e., infinitesimally small regions.
		A horizontal line is drawn at $10^{-3}$ to indicate the threshold below which the histogram is predominantly influenced by noise, suggesting caution in data interpretation within this range.
	}
	\label{fig:dataPanel}
\end{figure}

%%%%%%%%%%%%%%%%%%%%%%%%%%%%%%%%%%%%%%%%%%%%%%%%%%%%%%%%%%%%%%%%%%%%%%%%%%%%%%%%%%%%%%%%%%%
When exploring the distribution of axonal lengths in the brain, researchers frequently encounter two distinct types of data formats that exhibit significant differences.
In the ideal scenario, comprehensive information regarding the lengths of all neurons within the brain is available, as illustrated by the example of the Drosophila brain in the upper panel of Figure \ref{fig:dataPanel}a) \cite{CodexApp}.  
This dataset contains the axonal length of approximately thirty thousand neurons in the central connectome of Drosophila \cite{Scheffer2020}. 
Another similarly detailed dataset obtained with retrograde tracing provides the length for approximately two million axons crossing the white matter of the  mouse brain  \cite{Horvt2016}. 
However, in most cases, the data is coarse-grained, describing bundles of neurons that connect different regions of the brain \cite{Betzel2018}. 
In this scenario, we have an inter-region distance matrix, and numerous axons connecting the same two areas will be assigned equivalent lengths
 (depicted in the upper panel of Figure \ref{fig:dataPanel}b). 

Data concerning axonal distances were gathered for Drosophila, mouse, macaque, and human specimens. For mouse and Drosophila, we accessed both regional and axonal-level data, while for macaque and human, only regional-level data were available. Further information on the sources and formats of these data is provided below.

\textbf{Macaque (regional level)}
The macaque connectivity data at the regional level is derived from retrograde tract-tracing experiments as detailed in \cite{Markov2012}. This involved the use of fluorescent tracer injections in $28$ macaque monkeys to map the brain's connectivity. The resulting connectivity matrix has dimensions of $29\times91$, representing the connections weighted by neuron counts across $91$ cortical areas. For our analysis, the relevant data, including distances and connection weights, were specifically extracted from a $29\times29$ interareal subgraph of this matrix.

\textbf{Human (regional level)}
Human brain networks were reconstructed through deterministic tractography algorithms applied to diffusion-weighted MRI data.
The analyzed networks are composite representations of group-level connectivity patterns derived from individual subject networks (30 subjects).
The brain parcellation resolution spanned from 82 to 1014, with the selection of the smallest resolution, which closely approximated the segmentation size observed in the brains of other species (other resolutions were also utilized in the model evaluation phase).
The low-resolution parcellation comprises 68 cortical and 14 subcortical areas, whereas the high-resolution parcellation encompasses 1000 cortical and 14 subcortical areas.

For a more detailed description, refer to \cite{Betzel2018}.

\textbf{Drosophila (regional- and axonal level)}
The connectivity matrix of Drosophila was constructed using data from the FlyCircuit 1.1 database (http://www.flycircuit.tw), which contains images of 12,995 projections in the female Drosophila brain \cite{Chiang2011}.
In this study, 49 distinct brain segments were identified.
\newline
Furthermore, comprehensive neuron-level data for Drosophila was accessible through the flyWire database \cite{Scheffer2020}.
Our study utilizes the central connectome of the brain, which includes 32,422 neurons.

\textbf{Mouse (regional- and axonal level)}
Two region-level datasets derived from anterograde tracing were released nearly simultaneously in 2014 \cite{Oh2014,Zingg2014}. In their examination of the EDR model in mice, the authors of \cite{Horvt2016} combined these datasets to analyze the network structure. However, the determination of the EDR and its decay rate was based on retrograde tracing experiments, which accounted for the axonal lengths of approximately 2 million individual neurons \cite{Horvt2016} (as shown in Figure \ref{fig:dataPanel}a bottom panel). Subsequently, the same research group conducted another study on mice, extracting the regional network data from retrograde tracing experiments \cite{Gmnu2018}. This data was utilized to illustrate the region-level EDR in Figure \ref{fig:dataPanel}b bottom panel.

%%%%%%%%%%%%%%%%%%%%%%%%%%%%%%%%%%%%%%%%%%%%%%%%%%%%%%%%%%%%%%%%%%%%%%%%%%%%%%%%%%%%%%%%%%%
\section{Analytical model}
%%%%%%%%%%%%%%%%%%%%%%%%%%%%%%%%%%%%%%%%%%%%%%%%%%%%%%%%%%%%%%%%%%%%%%%%%%%%%%%%%%%%%%%%%%%
%\subsection{Analytical approach in one dimension}

\begin{figure}[!h]
      \centering
      \includegraphics[width=\linewidth]{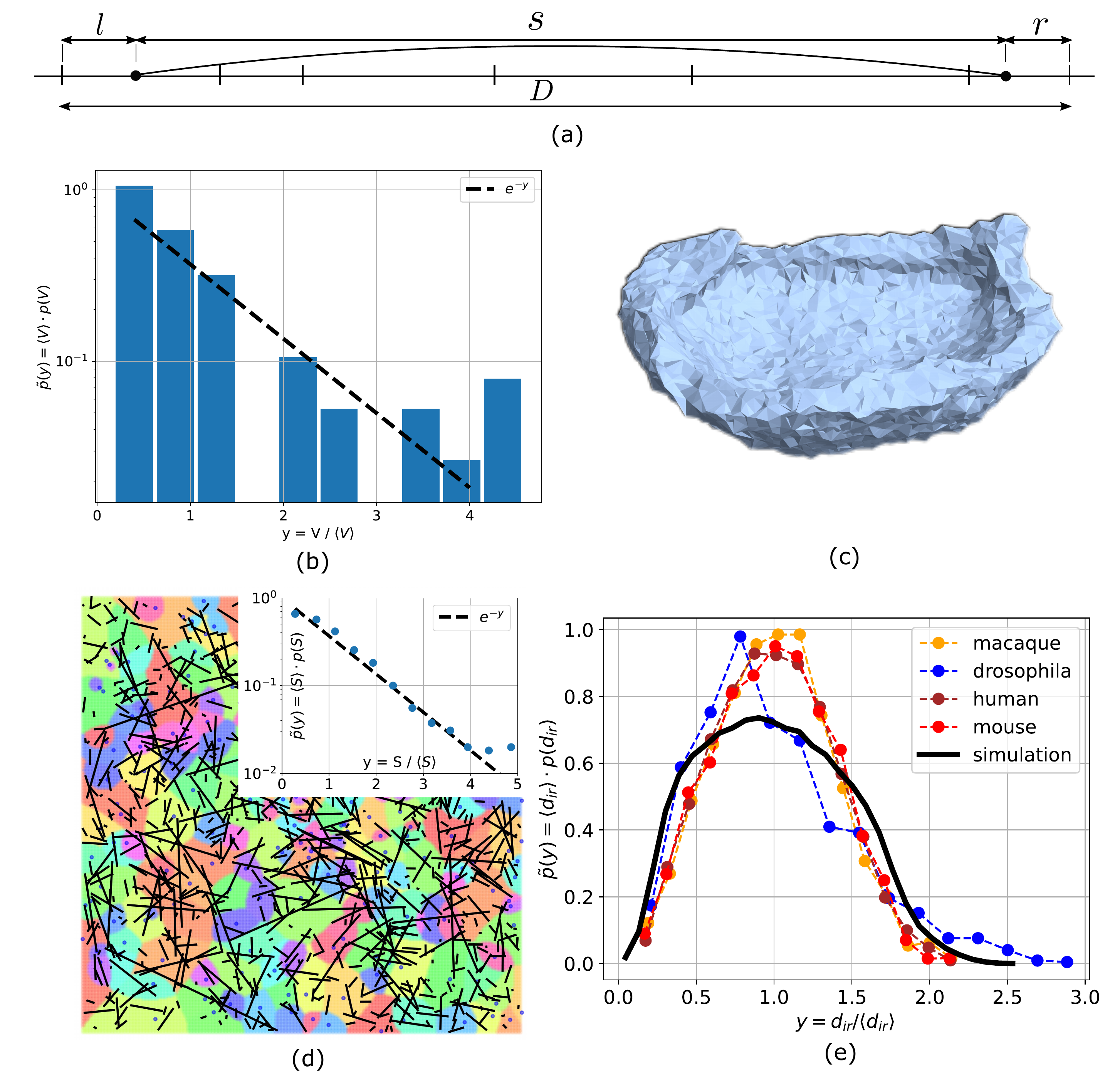}
      \caption{
      	Essentials of one and two-dimensional models of structural brain connectivity and data supporting the basic assumptions of these models. 
      	(a) Illustration depicting the distance coarse-graining process underlying the one-dimensional model. A  segment (axon) of length $s$ spanning across several cells (brain regions) will contribute to the connection length statistics with a length $D$ measured between the edges of the terminal cells (see Eqs. (\ref{E:conditional}, \ref{eq:oneDimModel}).
      	(b) Distribution of brain region volume sizes obtained from the mouse neurological null model \cite{Reimann2019}, providing support for the hypothesis of exponential region sizes in the model.
      	(c) A schematic representation of half of the mouse brain in three-dimensional space, emphasizing its two-dimensional character and minimal curvature (data sourced from \cite{Reimann2019}).
      	(d) Schematics of the computational experiment illustrating exponentially sized Voronoi cells and a sample of line segments randomly dropped on them (depicted by black lines).
      	The blue dots correspond to the randomly distributed seeds of the Voronoi cells. 
      	%The seeds crucial for calculating the coarse-grained distances.
      	The distribution of cell sizes follows a near-exponential trend, as depicted in the inset (the results presented here are derived from an ensemble of 30 averaged simulations).
      	(e) Simulation results on the distribution of distances between seed points compared with the distribution of distances between pairs of brain regions for the four datasets \cite{Markov2012, Betzel2018, Chiang2011, Gmnu2018}. In the two-dimensional Monte Carlo experiment, 30 seed points representing centers of brain regions were randomly placed. The experiment was repeated 1000 times to create an ensemble of distance matrices for robust statistical analysis.    	
      }
      \label{fig:methodology}
\end{figure}

To bridge the disparities between detailed neuron-level and broader region-level connectome data, we introduce a simplified one-dimensional model. 
In our model, we adopt two key assumptions: the validity of the EDR and, independently, an exponential distribution of brain region sizes. 
The latter hypothesis is distinct from the EDR and is based on the distribution of brain region volumes, which we derived by processing data generated from a biologically plausible null model of the mouse brain \cite{Reimann2019} (refer to Figure \ref{fig:methodology}b).
In this model, millions of neuron coordinates are provided each associated with a specific brain region.
We estimated the spatial extent within which neurons belonging to the same brain region are located, thereby determining the volume of these regions.

The model employs the infinite real axis as a scaffold, where brain regions, represented by non-overlapping intervals termed 'cells', partition the axis. 
Axons are conceptualized as line segments 'dropped' onto the axis to connect these cells. Each segment's endpoints directly associate with the cells they span, thereby simulating the connectivity between brain regions. 
Partitioning is achieved through the deployment of points distributed randomly along the axis, following a Poisson Point Process (PPP) with intensity parameter $\alpha$ inversely related to the mean cell size. This distribution process defines cell boundaries and results in cell sizes that adhere to an exponential distribution $\DistribAlphaBetaFunc_{\alpha}(x)$ similar to the one in Eq. (\ref{E:exponential}).
%, given by
%\begin{equation}\nonumber
%	\DistribAlphaBetaFunc_{\alpha}(x) = \alpha e^{-\alpha x}~.
%\end{equation}
To demonstrate this, consider the fundamental property of the PPP: 
the number of points in any interval of length $x$ follows a Poisson distribution with mean $\alpha x$. 
The probability of observing exactly zero points in an interval of length $x$ is given by 
$e^{-\alpha x}$, corresponding to the exponential distribution's survival function. Differentiating this function yields the PDF $\DistribAlphaBetaFunc_{\alpha}(x)$ confirming the exponential nature of cell size distribution within this model.

%Now, in line with the EDR, segments (connections) of length $s$ distributed according to $\DistribAlphaBetaFunc_{\lambda}(s)$ are placed randomly over the cells. 
%%This adherence with the EDR suggests that the distribution of axonal segment lengths follows an exponential pattern within the connectome.
%In this context, the symbol $\lambda$ represents the reciprocal of the mean axonal length.
%A connection of length $s$ can span a single or multiple cells, and the distance between the left boundary of the leftmost cell to the right boundary of the rightmost cell, denoted by $D$, represents the coarse-grained length under consideration.
%Formally, $D = s + x + y$, where $x$ and $y$ represent the distances of the connecting segment's endpoints to the left, and right margins, respectively (refer to Figure \ref{fig:methodology}a).
%The above definition of the coarse-grained connection length, $D$, is only preferred for the sake of analytical calculations. 
%While it might seem rather arbitrary, simulations show that  the outcome is qualitatively equivalent with more natural alternatives such as the distance between the centers of the left- and rightmost cells.
%By doing so, the conditional probability of coarse-grained distance, $D$, associated with a given segment length, $s$, can be expressed as:

In accordance with the EDR, connections of length \(s\), following the distribution \(\rho_{\lambda}(s)\), are randomly positioned across the cells, where \(\lambda\) signifies the reciprocal of the mean axonal length. 
A connection may span across one or several cells, with \(D\) representing the coarse-grained length of a connection, measured from the left boundary of the leftmost cell to the right boundary of the rightmost cell involved. 
Formally, \(D = s + l + r\), where \(l\) and \(r\) are the distances from the connecting segment’s endpoints to the respective left and right cell margins (see Figure \ref{fig:methodology}a).
This definition of \(D\) facilitates analytical calculations, although it may appear arbitrary. 
However, simulations demonstrate that this approach yields results qualitatively similar to those obtained using more intuitive measures, such as the distance between the centers of the terminal cells. 
The conditional probability of the coarse-grained distance, \(D\), for a specified segment length, \(s\), can be expressed as

%-
\begin{equation}\label{E:conditional}
	\DistribFunc(D|s) =\int_{\mathbb{R}^2_+} dl\,dr\, \DistribAlphaBetaFunc_\alpha(l)\DistribAlphaBetaFunc_\alpha(r)\delta(l + r + s - D)=
	\int_{\mathbb{R}_+} dl\,\DistribAlphaBetaFunc_\alpha(l) \DistribAlphaBetaFunc_\alpha(D - s - l)= \alpha^2(D-s)e^{-\alpha(D-s)}~.
\end{equation}
%-
Upon integrating over all possible segment lengths $s$ with pdf $\DistribAlphaBetaFunc_{\lambda}(s)$, we arrive at the distribution of coarse-grained distances:
%-
\begin{equation}
	\begin{aligned}
		\DistribFunc(D) &=\int_0^D ds\,\DistribFunc(D|s)\DistribAlphaBetaFunc_\lambda(s)= \alpha^2\lambda e^{-\alpha D}\int_0^D ds\,(D - s)e^{(\alpha-\lambda) s}=\alpha^2\lambda e^{-\lambda D}\int_0^D dq\,q e^{-(\alpha-\lambda) q} = \\
		&=\alpha^2\lambda D^2 e^{-\lambda D} \FFunc[(\alpha-\lambda) D]\ ,\quad \text{where} \quad \FFunc(z)\equiv\dfrac{1-e^{-z}(1+z)}{z^2}~,
	\end{aligned}
\label{eq:oneDimModel}
\end{equation}

with the mean of the distribution being:
\begin{equation}\label{eq:meaneq}
\langle D \rangle = \frac{\alpha+2\lambda}{\alpha \lambda}.
\end{equation}

\
%-

%Examining the limiting cases of this distribution function reveals intriguing properties.
In the limiting case of small regions and long connections ($\alpha \gg \lambda$), we get
\begin{equation}\nonumber
	\lim_{\alpha\rightarrow\infty} \DistribFunc(D)=\lambda e^{-\lambda D}~,
\end{equation}
which was to be expected  because the coarse-graining effect becomes negligible leading us back to the EDR rule, as depicted by the dashed line in Figure \ref{fig:dataPanel}b. 
In another extreme case, characterized by large regions and short connections ($\lambda \gg \alpha$), the result is the gamma distribution
\begin{equation}\nonumber
	\lim_{\lambda\rightarrow\infty}  \DistribFunc(D)=\alpha^2 De^{-\alpha D}~.
\end{equation}
In the scenario where regions and connections are of the same length scale ($\lambda \approx \alpha$), a gamma distribution is again observed:
\begin{equation}\nonumber
	\lim_{\lambda\rightarrow\alpha}  \DistribFunc(D)=\frac{1}{2}\alpha^3 D^2e^{-\alpha D}~.
\end{equation}
It is noteworthy that all three limiting cases of the two parameter distribution result in gamma functions with a single parameters.
This observation suggests that the general solution may also be readily approximated by a gamma distribution. 

A meaningful way for reducing the number of parameters is by rescaling the distribution to its mean, achieved by the transformation $D \rightarrow y={D}/{\langle D \rangle}$. 
In this way a comparison with the mean-rescaled experimental data will be straightforward.
If one eliminates $\alpha$, the rescaled distribution depending on the scale parameter of the connecting segments becomes
\begin{equation}\label{eq:normedFinal}
	 \langle D \rangle \DistribFunc(D) = \tilde{\DistribFunc}\left(y\right) = \frac{4 \lambda^3}{(\lambda-1)^2} y^2 e^{-\lambda y} \FFunc\left[\lambda\frac{3-\lambda}{\lambda-1}y\right]~,
\end{equation}
while eliminating $\lambda$ leads to
\begin{equation}\label{eq:normedFinalAlpha}
	\langle D \rangle \DistribFunc(D) = \tilde{\DistribFunc}\left(y\right) = \frac{\alpha^3}{\alpha-2} y^2  e^{-\frac{\alpha}{\alpha-2} y} \FFunc\left[\alpha\frac{\alpha-3}{\alpha-2}y\right]~.
\end{equation}

The one-dimensional coarse-graining model offers an excellent fit to the data (refer to Figure \ref{fig:result}a).
It's important to emphasize, that our focus here is on conducting a qualitative examination of the impact of coarse-graining, rather than undertaking a quantitative statistical analysis. Given the limited quality of the data (as it can be seen in Figure \ref{fig:dataPanel}b), and the simplicity of the model our approach did not prioritize achieving the most precise statistical fit.
Instead, we aimed to achieve a visually satisfactory fit.

The one-dimensional coarse-graining model demonstrates a strong alignment with the observed data, as evidenced in Figure \ref{fig:result}a. 
Here we wish to reiterate our emphasis on deriving qualitative insights into the effects of coarse-graining, rather than conducting a detailed quantitative statistical analysis. 
The quality of the data, illustrated in Figure \ref{fig:dataPanel}b, along with the simplicity of the model, guided our approach. 
This led us to prioritize achieving a visually satisfactory fit above rigorous statistical optimization.

To enhance visual clarity and highlight regions where data reliability is higher, we incorporated horizontal lines at $10^{-3}$ below which statistical errors become dominant.
A visually adequate fit can be attained by selecting the parameter $\lambda=2.3$ in Equation (\ref{eq:normedFinal}), corresponding to $\alpha=\frac{2\lambda}{\lambda-1}\simeq3.5$ after normalization to the mean  (setting $\langle D \rangle$ to $1$ in Eq. (\ref{eq:meaneq})).
The proximity of these values suggests that the model closely approximates a scenario where regions and connections share a similar length scale, leading to a gamma-like distribution function.

According to Equation (\ref{eq:normedFinalAlpha}) the model anticipates an overall non-monotonous dependence of the slope of the distribution on the resolution of the partitioning resulting in a U-shape, as illustrated in the bottom-left part of Figure \ref{fig:result}b.
However, in the $\alpha \in (3, \infty)$ interval the dependence is monotonous and shows a remarkable consistency with human brain data, as depicted in Figure \ref{fig:result}b.

%Conversely, in the $\alpha \in (3, \infty)$ interval, the slope decreases, resulting in a U-shape, as illustrated in the bottom-left part of Figure \ref{fig:result}b.
%Consistency with the behavior in the second regime is observed with data from various resolutions of the human brain, as depicted in Figure \ref{fig:result}b.increases within the interval $\alpha \in (2,3)$, the slope of the distribution function increases.
%
%Conversely, in the $\alpha \in (3, \infty)$ interval, the slope decreases, resulting in a U-shape, as illustrated in the bottom-left part of Figure \ref{fig:result}b.
%Consistency with the behavior in the second regime is observed with data from various resolutions of the human brain, as depicted in Figure \ref{fig:result}b.

%%%%%%%%%%%%%%%%%%%%%%%%%%%%%%%%%%%%%%%%%%%%%%%%%%%%%%%%%%%%%%%%%%%%%%%%%%%%%%%%%%%%%%%%%%%%%
\section{Simulations}
%%%%%%%%%%%%%%%%%%%%%%%%%%%%%%%%%%%%%%%%%%%%%%%%%%%%%%%%%%%%%%%%%%%%%%%%%%%%%%%%%%%%%%%%%%%%%
Our model  proposes that deviations from the exponential pattern observed in experimental data can be attributed to a simple region-level coarse-graining effect convolved with an underlying EDR. 
%The model has to key elements: the coarse-graining itself, i.e., the aggregation of axon endpoints and their association with the same region, and secondly, the one-dimensional framework allowing for an analytic approach. The latter is obviously a very gross approach and its errors might combine in a fortunate way with the coarse-graining hypothesis to provide the good fit with the data. 
The model incorporates two primary components: the coarse-graining process, which aggregates axon endpoints and associates them with the same region, and the one-dimensional framework that enables an analytical method. 
%This framework is a significant simplification, and we aim to ensure that the model's accuracy is genuinely due to the coarse-graining effect, rather than a fortuitous alignment of simplifications and assumptions.	
The latter, being a significant simplification, necessitates verification to ensure that the model's good fit with the data is indeed a result of the coarse-graining process, rather than a fortunate yet misleading alignment with the coarse-graining hypothesis.	
To that end, while maintaining the coarse-graining hypothesis, we modify some of the model's simplifying assumptions and even reconsider the validity of the EDR itself. 
This is undertaken at the expense of abandoning the analytical approach.

\subsection{One-dimensional models}
Monte-Carlo simulations of the model leading to Equations (\ref{eq:normedFinal}) and (\ref{eq:normedFinalAlpha}) give a perfect fit as shown in Figure \ref{fig:result}a (curve 13).
%The one-dimensional axis underwent segmentation by introducing random points to represent cell boundaries, resulting in an exponentially distributed distance between consecutive points (exponential cell size distribution).
%Subsequently, additional random points were distributed along this segmented line to denote the left boundaries of connections, or line segments.
%The length of each line segment was determined by an exponentially distributed number (in agreement with EDR), establishing the corresponding right boundary of the connection.
%The computation of the distance distribution for coarse-grained distances, specifically the distances between the nearest cell boundary on the left side of the line segment and the nearest cell boundary on the right side, exhibited a perfect match with the analytical solution.
%This congruence robustly supports the validity of the model.
Here the coarse-grained distance was calculated between the left boundary of the leftmost cell containing the left endpoint of the connecting segment to the right boundary of the rightmost cell including the right endpoint of the same "axon". 
Taking the distance between the geometric centers of the same cells produced a qualitatively similar distribution with some anomalies at short distances and relatively low resolutions. 
In this case short axon-level connections falling entirely within the same cell result in a vanishing coarse-grained distance. 
%Furthermore, when calculating the distances between the geometric centers of cells containing the left and right boundaries of the line segments placed on the one-dimensional axis, as opposed to the distances between the left and right boundaries of the same cells, the resultant distribution undergoes negligible alteration.
	For medium segment (axon) lengths, the shape of the distribution function resembles the one described by Eq. (\ref{eq:normedFinal}), albeit with a modified value for $\lambda$.

%-----------------------------------------------------------------------------------------
\subsection{Higher dimensions and curvature}

Another issue to address is the potential impact of dimensionality on simulation outcomes.
Figure \ref{fig:methodology}c provides a spatial representation of the half-brain of the mouse according to a neuron level null model \cite{Reimann2019}.
In this representation, neuron body coordinates were extracted from the data, and the corresponding surface, encompassing these, is depicted.
The figure reveals a cortex that is rather two dimensional than three. 
We expect this observation to stand also for the macaque and human brains and possibly to a lesser extent for the Drosophila. 
The influence of curvature is minimal at short connection distances, where the coarse-graining effect we investigate is most pronounced. 
Consequently, in this coarse-graining model, curvature is not expected to play a significant role, leading us to adopt a planar brain model.
Utilizing the same mouse brain model, we estimated the volume, and thus in a planar context, the area of various brain regions. 
Despite the limited number of regions leading to less robust statistics, the resulting distribution seems to follow an exponential pattern (see Figure \ref{fig:methodology}b).
%Considering the relatively planar structure of the mouse brain and the limited significance of curvature in the modeling process, no additional simulations were conducted in three dimensions.
We conducted two-dimensional simulations generating custom Voronoi tesselations by assigning exponential weights to the seeds of 200 cells. 
The weighting scheme influenced the distribution of cell sizes, closely aligning them with an exponential distribution, as depicted in the inset of Figure \ref{fig:methodology}d.
The previously applied EDR together with the coarse-graining procedure was repeated by randomly placing connections of exponentially distributed lengths over the partitioned plane.
The results of the simulation experiment demonstrate a remarkably close resemblance to both the one-dimensional model and the data, as illustrated in Figure \ref{fig:result}a.
These results support the conclusion that dimensionality does not qualitatively impact the output of the coarse-grained EDR model.
%Furthermore, the two-dimensional model can effectively capture the peaked distribution observed in the distances between all pairs of brain regions.
Figure \ref{fig:methodology}e showcases the distribution of distances between seeds in the Voronoi-partitioned two-dimensional space, revealing a peaked distribution curve.
This similarity to empirical data supports the realism of the two-dimensional model. Importantly, the choice to use geometric centers of cells rather than the seeds for estimating inter-region distances does not significantly alter the observed distributions, whether for coarse-grained 'axonal' distances or for inter-region distances
%the resulting distribution remains largely unaltered when considering the distances between the geometric center of exponentially sized Voronoi cells.

%-----------------------------------------------------------------------------------------
\subsection{EDR, sufficient or necessary?}

\begin{figure}[h!]
	\centering
	\includegraphics[width=\linewidth]{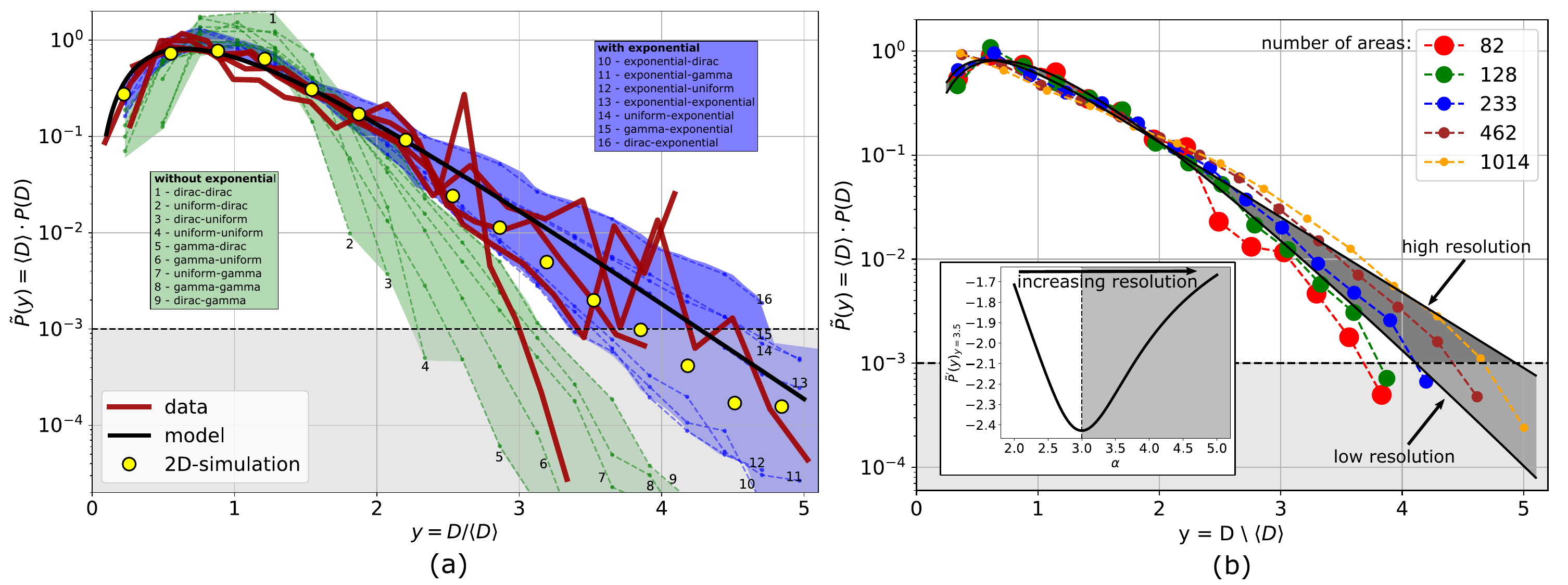}
	\caption{
		Comparative analysis of analytic calculations and simulations against empirical data.
		(a) Cumulative findings from analytical modeling, computational simulations - including a two-dimensional experiment and varied permutations of axonal length and spatial segmentation distributions - and comparison with datasets.
		The same weighted intra-cortical connection distances that are detailed in Figure \ref{fig:dataPanel}b \cite{Markov2012, Betzel2018, Chiang2011, Gmnu2018} are represented by red lines, while the analytical solution, expressed by Equation (\ref{eq:normedFinal}) with $\lambda=2.3$, is depicted by a black line.
		Yellow dots represent the simulation results from the two-dimensional experiment, featuring exponentially sized Voronoi cells.
		The dashed-dotted lines represent outcomes from one-dimensional simulations that were conducted using various pairs of distributions for axonal lengths and brain region sizes. Specifically, the blue region is associated with simulation scenarios where at least one of the two distributions—either the axon length or the brain region size—is exponential. Conversely, the green regions denote scenarios where neither distribution is exponential.
		All simulations were conducted considering $\lambda=2.3$ for the inverse of the mean of the distribution of axon lengths (and with the corresponding $\alpha=\frac{2\lambda}{\lambda-1}$ derived from the condition of the mean, $\langle D\rangle$, equal to one in the distribution coming from Eqs. (\ref{eq:oneDimModel}, \ref{eq:meaneq}).
		(b) The analytical solution's boundaries within the range $\alpha \in (3, 5)$, as expressed in Equation (\ref{eq:normedFinalAlpha}), are evaluated in connection with human brain data \cite{Betzel2018}.
		Various resolutions of brain segmentation, spanning from 82 to 1014 areas, are considered.
		The observed impact of changes in resolution on the distribution function consistently aligns with the predictions of the model within this interval.
		This interval corresponds to the right side of the U-shaped function depicting the slope of the distribution function in Equation (\ref{eq:normedFinalAlpha}).
		The slope calculated at $y=3.5$ is also illustrated as a function of the $\alpha$ parameter in the lower-left part of the figure.
			Note that the red and green curves falling outside the theoretical limit correspond to cases where the number of subcortical regions (14) is relatively high ($>10\%$) compared to cortical regions, indicating that our model is more suitable for analyzing cortical-level neuron data.		}
	\label{fig:result}
\end{figure}

To further validate the EDR's universality, it is insufficient to rely on a model that merely predicts lower probability density function values at extremely short and long distances. Instead, it is crucial to demonstrate that the coarse-graining process inherently necessitates an exponential input to accurately reflect empirical data. 
Consequently, we conducted additional simulations employing alternative distributions such as Dirac-delta, uniform, Tsallis-Pareto, or gamma, in place of exponential distributions for both cell size and connection length distributions. 
Parameters for these distributions were adjusted to match the $\alpha$ and $\lambda$ parameters that yielded the optimal fit shown in Figure \ref{fig:result}a ($\lambda=2.3$ and $\alpha=\frac{2\lambda}{\lambda-1}\simeq3.5$). The mean-normalized gamma distribution's exponent was set to 3, though our findings suggest that variations in this exponent do not significantly affect the conclusions. 
As depicted in Figure \ref{fig:result}a, the implementation of an exponential distribution for either connection lengths or cell sizes proves sufficient for replicating the observed coarse-grained data.

%%%%%%%%%%%%%%%%%%%%%%%%%%%%%%%%%%%%%%%%%%%%%%%%%%%%%%%%%%%%%%%%%%%%%%%%%%%%%%%%%%%%%%%%%%%
\section{Conclusions}
%%%%%%%%%%%%%%%%%%%%%%%%%%%%%%%%%%%%%%%%%%%%%%%%%%%%%%%%%%%%%%%%%%%%%%%%%%%%%%%%%%%%%%%%%%%

We conducted an extensive investigation into the Exponential Distance Rule (EDR)'s applicability across diverse datasets, focusing on detailed neuron-level connection data for mice and Drosophila. 
Our analysis reveals that the length distribution of axonal connections closely adheres to the EDR. 
Additionally, we examined region-level brain connectivity data for Drosophila, mice, macaques, and humans. 
We interpret these region-level connections, once weighted by the number of axons involved, as a coarse-grained depiction of the neuron-level connectome. 
Our findings indicate that the weighted length distributions of these region-level connections, upon normalization to the mean axonal lengths, align with a universal master curve across all studied species.
 Notably, this master curve deviates from a purely exponential form, instead bearing resemblance to a gamma distribution.
%It was revealed that weighted (regional) intercortical data coarse-grains the distribution of neuron lengths, resulting in a pattern that is not purely exponential but instead resembles a gamma distribution.
%This provides compelling evidence for the universality of these observations in this context.
We demonstrated that deviations from the EDR can be attributed to the coarse-graining effect observed in region-level connections, with the master curve's shape being analytically and simulationally explained. 
The simplicity of our one-dimensional model prompted us to test its boundaries, including considerations of dimensionality and spatial curvature. 
Through specialized Voronoi tessellation on a two-dimensional plane, featuring an exponential distribution of cell sizes, we observed that the variation in coarse-grained distances closely mirrors the one-dimensional scenario. 
This affirms the Exponential Distance Rule (EDR)'s role as a sufficient condition for aligning with all experimental data. 
Further, we probed the necessity of the EDR for a model's consistency with empirical observations. 
Evaluating models that utilize non-exponential distributions for both axon lengths and brain region sizes, we concluded that an exponential distribution is critical for accurately representing either the length of connections (axons) or the size of regions (cells).
%The results suggest that the model performs well within these boundaries.
The validation of our simple model is further reinforced by the observation that variations in the master curve, resulting from different brain partitioning resolutions—namely, the number of disjoint regions—align with the model’s predictions. 
%Consequently, our findings imply that the EDR can be applied more broadly, extending its relevance not only to rodents (mice) and primates (macaques) but potentially also to humans and other species for which exhaustive axon level data are not available.
Thus, our findings indicate that the EDR's relevance is not confined to specific species such as rodents (mice) and primates (macaques), but may also extend to humans and other species for which detailed axon-level data are not available.

Further research should consider the prediction of the decay parameter $\lambda$ for species without detailed tract-tracing data. This could leverage existing scaling laws for brain parameters such as white matter volume, gray matter volume, and neuron count \cite{Liu2003,HerculanoHouzel2007,HerculanoHouzel2010,VenturaAntunes2013,HerculanoHouzel2012}. Given the inverse relationship between $\lambda$ and average axonal length, and its expected correlation with white matter volume, this approach may yield new scaling laws. The EDR-based random network model's success in mimicking key features of cortical networks \cite{ErcseyRavasz2013} suggests the utility of this method. Predicting $\lambda$ values and network characteristics for unexplored species could enhance comparative studies of neural architectures.

%Subsequent investigations could explore the prospect of predicting the decay parameter $\lambda$ for other species where costly tracing experiments have not yet been conducted.
%This prediction could rely on established scaling laws governing various brain parameters, such as the volume of white matter, gray matter, and the number of neurons \cite{Liu2003,HerculanoHouzel2007,HerculanoHouzel2010,VenturaAntunes2013,HerculanoHouzel2012}.
%Given that the decay rate is inversely related to the average axonal length, a direct correlation with the volume of white matter is anticipated, offering potential scaling laws to explore.
%The random network model based on the EDR has demonstrated the ability to replicate numerous intriguing properties of the inter-regional cortical network \cite{ErcseyRavasz2013}.
%Predicting the $\lambda$ rates and network properties for other species could present an intriguing avenue for future research endeavors.

%%%%%%%%%%%%%%%%%%%%%%%%%%%%%%%%%%%%%%%%%%%%%%%%%%%%%%%%%%%%%%%%%%%%%%%%%%%%%%%%%%%%%%%%%%%
\section{Acknowledgements}
%%%%%%%%%%%%%%%%%%%%%%%%%%%%%%%%%%%%%%%%%%%%%%%%%%%%%%%%%%%%%%%%%%%%%%%%%%%%%%%%%%%%%%%%%%%
Work supported by the Romanian CNCS-UEFISCDI grants PN-III-P4-ID-PCE-2020-0647 (Zs.I.L.),  PN-III-P4-ID-PCE-2021-0408 (M.E.-R.), COFUND-FLAGERA-ModelDXConsciusness (M. E.-R., Zs.I.L.) and ERANET-NEURON-2-UnscrAMBLY (M.E.-R.). The work of M.J. and Zs.I. L. is also supported by the Collegium Talentum Program of Hungary.

\printbibliography

\end{document}